\documentclass{sig-alternate}
\usepackage{mathptmx} 

\usepackage{fancyhdr}
\usepackage[normalem]{ulem}
\usepackage[hyphens]{url}
\usepackage[sort,nocompress]{cite}
\usepackage[final]{microtype}
\usepackage[keeplastbox]{flushend}
\usepackage[bookmarks=true,breaklinks=true,letterpaper=true,colorlinks,citecolor=blue,linkcolor=blue,urlcolor=blue]{hyperref}

\fancypagestyle{firstpage}{
  \fancyhf{}
  
  \fancyhead[C]{\vspace{10pt}\normalsize{This is an expanded iteration of our previously published paper \cite{KlimisICSE24}, featuring additional elaboration and insights.}\\\vspace{-25pt}} 
  \fancyfoot[C]{\thepage}
}

\title{Lost in Interpretation: Navigating Challenges in Validating Persistency Models Amid Vague Specs and Stubborn Machines, All with a Sense of Humour} 

\author{
  \Large Vasileios Klimis \\
  \large Queen Mary University of London \\
  \small \texttt{v.klimis@qmul.ac.uk}
  \and
  \Large Alastair F. Donaldson \\
  \large Imperial College London \\
  \small \texttt{alastair.donaldson@imperial.ac.uk}
  \and
  \Large Viktor Vafeiadis \\
  \large MPI-SWS \\
  \small \texttt{viktor@mpi-sws.org}
  \and
  \Large John Wickerson \\
  \large Imperial College London \\
  \small \texttt{j.wickerson@imperial.ac.uk}
  \and
  \Large Azalea Raad \\
  \large Imperial College London \\
  \small \texttt{azalea.raad@imperial.ac.uk}
}

\usepackage{defs}

\begin{document}
\maketitle
\thispagestyle{firstpage}
\pagestyle{plain}


\begin{abstract}
Memory persistency models provide a foundation for persistent programming by specifying which (and when) writes to non-volatile memory (NVM) become persistent. 
Memory persistency models for the Intel-x86 and Arm architectures have been formalised, but
not empirically validated against real machines. 
Traditional validation methods 
used for memory \emph{consistency} models 
do not straightforwardly apply because a test program cannot directly observe when its data has become persistent: it cannot distinguish between reading data from a volatile cache and from NVM. 
We investigate addressing this challenge using a commercial off-the-shelf device that intercepts data on the memory bus and logs all writes in the order they reach the memory.
Using this technique we conducted a litmus-testing campaign aimed at empirically validating the persistency guarantees of Intel-x86 and Arm machines.
%
We observed writes propagating to memory out of order, and took steps to build confidence that these observations were not merely artefacts of our testing setup. 
However, despite gaining high confidence in the trustworthiness of our observation method, our conclusions remain largely negative. 
We found that the Intel-x86 architecture is not amenable to our approach, and on consulting Intel engineers discovered that there are currently no reliable methods of validating their persistency guarantees. 
For Arm, we found that even a machine recommended to us by a persistency expert at Arm did not match the formal Arm persistency model, due to a loophole in the specification.
Nevertheless, our investigation and results provide confidence that if Intel were to produce machines with more transparent persistency behaviour, or if Arm machines with proper persistency support were to become available, our approach would be valuable for empirically validating them against their specifications.

\end{abstract}

\section{Introduction}\label{sec:introduction}


Non-volatile memory (NVM) technologies such as Samsung's recently announced
Memory-Semantic SSD~\cite{samsungpressrelease} and Kioxia's
XL-FLASH~\cite{techinsightsblog}
simultaneously provide
(1) low-latency, high-throughput, and fine-grained data transfer capabilities as DRAM does; and
(2) durable and high-capacity storage as SSD does.
As such, NVM (\aka persistent memory) has the potential to radically change the way we build fault-tolerant systems
by optimising traditional and distributed file systems~\cite{strata,winefs,kuco,octopus,linefs,nova}, transaction processing systems for high-velocity real-time data~\cite{sstore}, distributed stream processing systems~\cite{kafka}, and stateful applications organised as a pipeline of cloud serverless functions interacting with cloud storage systems~\cite{olive,beldi}.

To provide a rigorous foundation for reasoning about persistent programs (operating over NVM), several formal models of memory \emph{persistency} have been proposed for existing hardware architectures. These include the Px86 model \cite{RaadWNV20} for the Intel-x86 architecture and the PArm model \cite{RaadWV19} for the Arm architecture.
These models have been developed in collaboration with architects and engineers from the respective vendors, and they have been judged to align to the prose text of the respective architectural specifications. However, they have not been empirically validated against real machines.

Empirical validation is critical for two key reasons. First, it is crucial to hold hardware vendors accountable by testing whether their hardware implementations
in deployed machines conform to their specifications (this has not always been the case when validating memory \emph{consistency} models~\cite{AlglaveMSS10, AlglaveBDGKPSW15}).
Second, it is important to confirm that the persistency specifications are accurate. Specifically, our primary aim is to ascertain that they are \emph{sound}, in that they do not forbid behaviours that can indeed happen on real machines. It is also desirable that persistency specifications are not unduly weak, in that they do not allow too many behaviours that cannot be observed on real machines.

In the context of memory \emph{consistency} models, empirical validation has become commonplace. Tools such as \litmus~\cite{litmus} have been widely used to run large numbers of small multi-threaded test cases (called \emph{litmus tests}) on machines-under-test, in order to see which threads can see which writes in which order.
However, in the context of memory \emph{persistency}, the situation is more complicated. A \litmus-like approach will not work because a running program cannot directly observe when its data has become persistent. Specifically, there is no way for it to reliably determine if it is reading values from a volatile cache or from persistent memory. 

A naive approach would be to contrive a crash (\eg turn off the machine) while the program is running and then read persistent data from the NVM once the machine restarts. However, this only allows one to observe the latest persisted write for each memory location, and is not sufficient to infer the order in which earlier writes persisted. Moreover, continually power-cycling the machine-under-test makes it infeasible to run the large number of tests required for high-coverage validation, and in any case, it would be nearly impossible to schedule the `crashes' precisely enough. 

\begin{figure}
    \centering
    \includegraphics[width=\columnwidth]{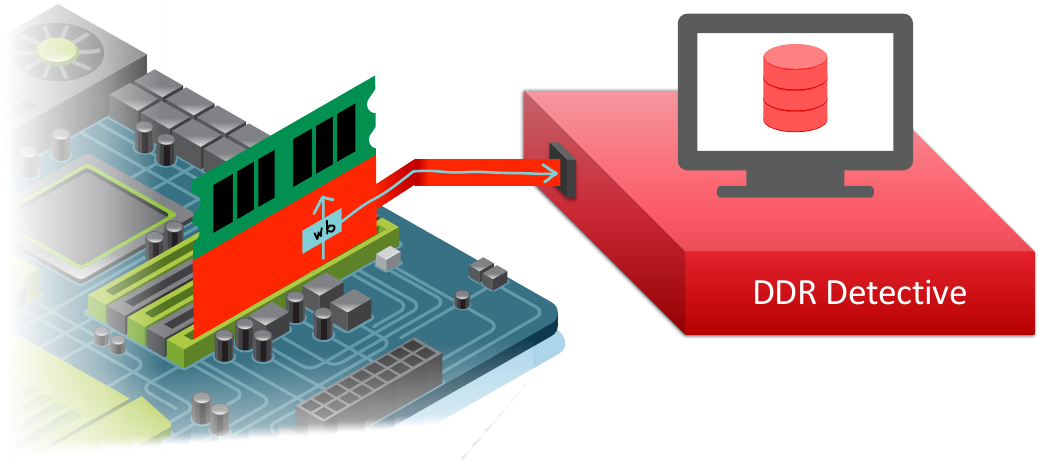}
    \caption{In our empirical setup the interposer sits between the DIMM and the DIMM socket on the motherboard, and the attached \DDRDetective{} logs specified memory accesses.}
    \label{fig:setup}
\end{figure}

This leads us to the approach we put forward in this paper: to observe writes as they reach the NVM, using a special interposer that sits between the motherboard and the memory module and sends data to a device called the \DDRDetective{}~\cite{FuturePlus} (see \cref{fig:setup}). By using the \DDRDetective{} to record every memory access while a litmus test is running and then inspecting the log, we should be able to deduce the order in which writes become persistent.

To make this approach work, a key challenge we needed to overcome was to decipher the mapping between the virtual memory addresses used in our litmus tests and the DRAM addresses that travel along the memory bus.
We could not simply write distinctive values like {\tt 0xDEADBEEF} and look for them in the log: the \DDRDetective{} only records the memory addresses that are accessed and not the values that are written. Instead, our solution involves performing a fixed number of writes to each memory location used in the litmus test -- e.g.\ 100 writes to {\tt x} and 200 writes to {\tt y} -- and then searching the log for the two locations appearing \emph{exactly} those many times (in the rare case that this holds for multiple locations the attempt can simply be repeated).

We have applied our approach to an Intel-x86 machine and an Arm machine, and have been able to confirm using simple litmus tests that writes can indeed enter memory in an out-of-order fashion (\ie disagreeing with the order prescribed by the formal persistency models), even when they are `persist-separated'; that is, separated by explicit (sequences of) instructions that a programmer can use to persist writes in a certain order. Nevertheless, our results remain inconclusive because of various subtleties in both architectures.

\begin{figure}
    \centering
    \includegraphics[width=\columnwidth]{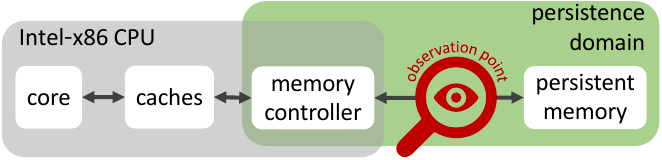}
    \caption{Persistency in a typical Intel-x86 system}
    \label{fig:x86_system}
\end{figure}

In the case of the Intel-x86 machine, the difficulty is that data headed to memory passes first through a battery-backed \emph{write-pending queue} (WPQ), which is part of the memory controller on the processor~\cite{Scargall2020}. This means that data becomes persistent before it has left the processor, and hence any observations we make between the processor and the memory module are too late down the pipeline and are immaterial to the persist order (see \cref{fig:x86_system}).
\begin{figure}
    \centering
    \includegraphics[width=\columnwidth]{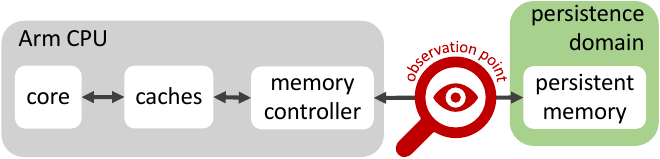}
    \caption{Persistency in a typical Arm system}
    \label{fig:arm_system}
\end{figure}

Nonetheless, we applied our approach to an Intel-x86 machine to see whether the WPQ does in fact reorder writes. We were able to observe, via simple litmus tests, that data does indeed propagate from the processor to memory in an out-of-order fashion.
Although (due to the battery-backed WPQ) this observation has no bearing on whether or not the Intel-x86 machine conforms to the Px86 persistency model~\cite{RaadWNV20}, 
our experiments lead to two significant observations. 
First, our results may be relevant in the context of \emph{remote direct memory accesses} (RDMA). 
Specifically, RDMA facilitates communication between two machines over a network by allowing each machine to directly access the (remote) memory of another through the \emph{network interface card} (NIC) technology, thus completely bypassing the operating system and the CPU (and its WPQ) of both machines, leading to low latencies and high bandwidth.
That is, the persistent memory locations of a machine can be accessed both locally (by the CPU) and remotely (by the NIC), 
As such, when the same locations are written to both locally and remotely, one can no longer rely on the WPQ to persist pending writes to memory as remote writes bypass the WPQ.


Second, our results for Intel-x86 highlight an important shortcoming, namely the lack of transparency in Intel CPU designs: it should be possible for Intel's end-users to empirically validate the CPUs in deployed machines, to hold vendors to account. As is, there is currently no feasible and reliable way to conduct such empirical validation \emph{independently} -- we have confirmed this with one of the principal persistency engineers at Intel.
To our knowledge, ours is the first work to shed light on this important shortcoming.

In the case of Arm machines, there is no WPQ-like battery-backed component and the setup is as in \cref{fig:arm_system}. This means that on a machine with support for persistent memory, the point at which the DDR Detective intercepts memory writes \emph{should} be the correct point at which to potentially observe reorderings that violate the Arm persistency model.
To investigate this, we procured a machine that was recommended to us for this project by a persistency expert at Arm -- a machine that is fully compliant with the Arm v8.2 architecture,
which was announced in 2016 and includes primitives for persistent programming~\cite{arm_persistency_announcement}.
However, it turned out that (despite the expert recommendation) this system does not support NVM, and a loophole in the Arm user manual allows it to avoid actually guaranteeing a point-of-persistency (at which the writes must become persistent). As such, even though our experimental setup was able to observe persist-separated writes propagating to NVM out-of-order, the loophole means that this does not technically violate the specification.
On further investigation, it appears that there is at present no commercially available Arm system that supports NVM.
Nonetheless, our empirical efforts on Arm led to two main results. 
First, they highlight a significant gap in understanding and communicating the persistency guarantees of Arm machines. After all, if an Arm persistency expert mistakenly assumes that a given Arm machine provides certain persistency guarantees, then there is little chance for the end-users/programmers of this machine to arrive at correct assumptions. 
Second, if an Arm system with persistency support were to come to market, our approach would be effective for validating its persistency claims.

\paragraph{Contributions} We claim two research contributions:
\begin{enumerate}
\item an experimental setup demonstrating the feasibility of validating memory persistency models such as Px86~\cite{RaadWNV20} and PArm~\cite{RaadWV19}; and
\item empirical evidence that both Intel-x86 and Arm machines allow memory accesses to be reordered between the CPU and the memory, which may have implications for using these machines with RDMA technologies.
\end{enumerate}

\paragraph{Supplementary material} Our approach relies on a specialised analysis tool, so reproducing our results in full is likely to be challenging. However, to increase transparency in our methodology, our automation scripts, litmus tests, and raw data are accessible as supplemental material at: \url{https://zenodo.org/records/10427213}.

\paragraph{Outline} The remainder of this article is organised as follows. In \cref{sec:background} we provide the necessary background on persistency in Arm and Intel-x86 architectures. In \cref{sec:infrastructure}, we describe the empirical testing infrastructure, paying particular attention to the experimental setup and the challenges it presents. In \cref{sec:evaluation} we present an experimental assessment and discussion of the findings. We discuss related work in \cref{sec:related} and conclude in \cref{sec:conclusion}.

\section{Background}\label{sec:background}
Non-volatile memory (NVM) retains its contents in the case of a crash, \eg due to power loss. As with volatile memory, NVM is available as a dual in-line memory module (DIMM) package that can be attached to the memory bus of a motherboard. Modern architectures such as Intel-x86 and Arm provide instructions for wrting back cache contents to memory; these can be used by expert programmers to control the order in which writes to NVM become persistent. 


\begin{figure}[t!]

    \begin{minipage}{0.49\columnwidth}
      \begin{litmustest}[H]
    
    \caption{\scshape Intel} \label{alg:Intel}
    
    \vspace*{.2cm}
      \end{litmustest}
    \end{minipage}
       \hfill
    \begin{minipage}{0.49\columnwidth}
      \begin{litmustest}[H]
    
    \caption{\scshape Arm} \label{alg:Arm}
    
    \vspace*{.2cm}
      \end{litmustest}
    \end{minipage}

    \begin{minipage}{\columnwidth}
    \vspace*{.2cm}
    \colorbox{lightgreen}
    {
        \begin{minipage}{0.44\columnwidth}
        
          \begin{litmustest}[H]
        \SetAlgoLined
        \DontPrintSemicolon
        
        \textcolor{gray}{\nl} $x \gets 1$\;
        \textcolor{gray}{\nl} $ \mathit{clflush(x)} $\;
        \textcolor{gray}{\nl} $y \gets 1$\;
        \;
          \end{litmustest}
        \end{minipage}
    }
       \hfill
    \colorbox{lightgreen}
    {
        \begin{minipage}{0.44\columnwidth}
        
          \begin{litmustest}[H]
          \SetAlFnt{\small}
        \SetAlgoLined
        \DontPrintSemicolon
        
        \textcolor{gray}{\nl} $x \gets 1$\;
        \textcolor{gray}{\nl} $ \mathit{dc\_cvap(x)} $\;
        \textcolor{gray}{\nl} $ \mathit{dsb(sy)} $\;
        \textcolor{gray}{\nl} $y \gets 1$\;
        
          \end{litmustest}
        \end{minipage}
    }

        \vspace*{.1cm}

\colorbox{CanaryYellow}{
    \parbox{.955\columnwidth}{\centering
                    {\tt upon recovery:} $~~y=1 \implies x=1$
    }
}

    \end{minipage}
\end{figure}

\paragraph{Persistency in the Intel-x86 architecture}
The persistency semantics of the Intel-x86 architecture has been formalised as the Px86 model by Raad \etal \cite{RaadWNV20,ChoLRK21,RaadMV22}. Px86 describes a \emph{relaxed} persistency model, in that the \emph{coherence order} (\ie the order in which writes are made visible to other threads) and the \emph{persist order} (\ie the order in which writes are persisted to memory) may disagree. 
To see this, consider the simple sequential program $x \gets 1; y \gets 1$, where we write $1$ to $x$ and then write $1$ to $y$. 
If a crash occurs during the execution of this program, at crash time either write may or may
not have already persisted to NVM and thus $x, y \in \{0, 1\}$ upon recovery. 
However, the relaxed nature of Px86 allows for a surprising behaviour that is not possible during normal
(non-crashing) executions: at no point during the normal execution of this program is the state
$x=0, y=1$ observable, because the two writes cannot be reordered under Intel-x86. 
Nevertheless, in case of a crash it \emph{is} possible to observe $x=0, y=1$ after recovery under Px86. 
This is due to the relaxed persistency of Px86: the coherence order ($x$ before $y$), is different from the persist order ($y$ before $x$). 

To allow more control over when and how writes are persisted to NVM, Intel-x86 provides explicit
\emph{persist} instructions such as $\mathit{clflush(x)}$, as illustrated in \cref{alg:Intel}: executing $\mathit{clflush(x)}$ persists the earlier write on $x$ (\ie $x \gets 1$) to memory. 
As such, if a crash occurs during the execution of this program and upon recovery $y=1$ is observed,
then $x=1$ is guaranteed to also be observed. 
That is, if $y \gets 1 $ has executed and persisted before the crash, then this must also be the case for the earlier
$x \gets 1$, due to the intervening $\mathit{clflush(x)}$.

\paragraph{Persistency in the Arm architecture}
As with Intel-x86, the Arm architecture follows a relaxed persistency model~\cite{RaadWV19,ChoLRK21} and provides instructions for persistent programming. 
Specifically, the $\mathit{dc\_cvap(x)}$ is the Arm analogue of $\mathit{clflush(x)}$ on Intel-x86 and ensures that earlier writes on $x$ are propagated to the \emph{point of persistence} (PoP), 
\ie that they have reached the `persistence domain' and will persist beyond subsequent power failures. 
However, while $\mathit{clflush(x)}$ is \emph{synchronous} (\ie it blocks execution until pending writes are persisted), $\mathit{dc\_cvap(x)}$ is \emph{asynchronous} (\ie execution is not stalled and the writes will be persisted at some future point). 
To remedy this, one must follow $\mathit{dc\_cvap(x)}$ with a \emph{full data synchronisation barrier}, $\mathit{dsb(sy)}$, which blocks until the effects of all earlier (in program order) $\mathit{dc\_cvap}$ instructions take place. 
Together, $\mathit{dc\_cvap(x)}$ and $\mathit{dsb(sy)}$ form an explicit `persist': a sequence of instructions that, if placed between two writes, will ensure that those writes become persistent in program order.
This is illustrated in \cref{alg:Arm}: executing the persist sequence $\mathit{dc\_cvap(x); dsb(sy)}$ persists the earlier write on $x$ (\ie $x \gets 1$).
As such, if a crash occurs during the execution of this program and upon recovery $y=1$ is observed then $x=1$ is also guaranteed to be observed.

The typical persistence domain in an Arm system comprises just the persistent memory itself \cite[p.~7]{wang22}. This means that if we can observe the order of writes as they pass along the memory bus (see \cref{fig:arm_system}), we can deduce the order in which they persist.

%

\SetKwComment{tcp}{\textcolor{gray}{//} }{}%

\begin{figure*}
\captionsetup{width=1\textwidth}
\includegraphics[width=1\textwidth]{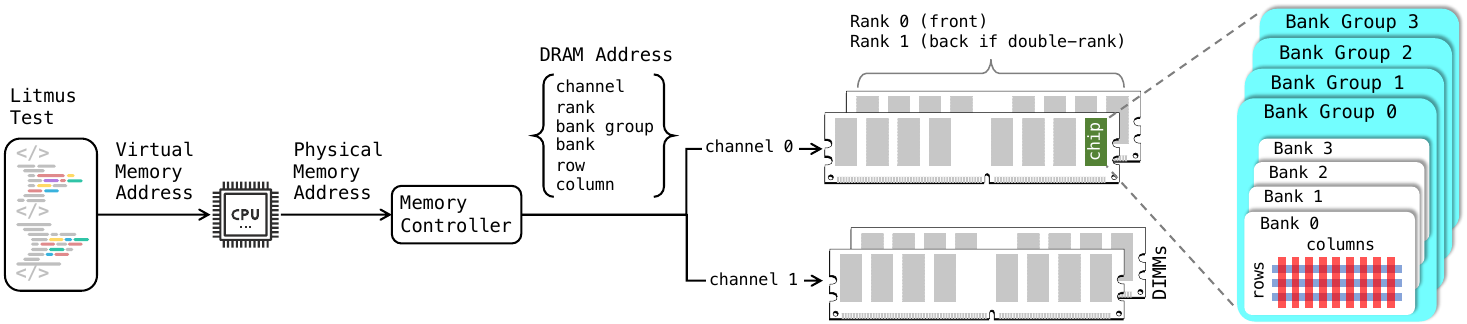}
  \caption{DRAM addressing and spatial organisation}
  \label{fig:DRAMorganisation}
\end{figure*}

\paragraph{Intel-x86 versus Arm}
A key difference between an Intel-x86 and a typical Arm setup is that the persistence domain of Intel-x86 systems includes the \emph{memory controller} (MC), which is battery-backed and thus has sufficient power to flush pending writes to memory in case of a power failure, using a mechanism called \emph{asynchronous DRAM refresh} (ADR)~\cite[p.~17]{Scargall2020}. 
That is, our observation point on the memory bus (see \cref{fig:x86_system}) is not in the correct place -- the order in which writes travel along the memory bus is not necessarily the same as the order in which they reached the MC and were thus persisted.
That is, the correct point for observing the persist order would be as they arrive in the MC. 
As our conversations with Intel engineers have revealed, we are not aware of a reliable mechanism for directly observing the order in which writes enter the MC.
Moreover, recent Intel-x86 systems provide \emph{enhanced ADR} (eADR)~\cite[p.~18]{Scargall2020}, extending the persistence domain to include the L2 and L3 caches, thus moving the correct observation point for the persist order still further from the memory bus. 

\paragraph{Virtual, physical, and geometric memory addresses}
Our litmus tests refer to virtual memory addresses ($x$ and $y$). It is straightforward to determine in software how these virtual addresses map to physical memory addresses. 
However, the addresses of the writes that travel on the memory bus and are intercepted by the \DDRDetective{} follow a 'geometric' structure: they refer to the internal organisation of the memory chip, comprising rank, bank, row, and column numbers, as illustrated in  \cref{fig:DRAMorganisation}.

The mapping between physical addresses and geometric addresses is handled by the CPU's memory controller. The mapping is rarely straightforward, for several reasons. 
The memory controller tries to choose a mapping that maximises the number of memory accesses that can be serviced in parallel by balancing traffic among different memory modules and minimising bank conflicts. It may also obfuscate the mapping for security reasons, \eg to counteract RowHammer attacks~\cite{6853210} (where a row of memory cells can be corrupted by repeated writes to geometrically adjacent rows), and it may perform wear-levelling when targeting certain solid-state memories. 
In addition, DRAM modules are known to internally rearrange rows \cite{10.1007/978-3-030-00470-5_3}, thereby adding further complexity to the mapping process.
The details of the mapping are often considered proprietary and kept as trade secrets, resulting in a lack of documentation and information available. 
As such, automatically determining which geometric addresses correspond to the virtual/physical addresses used by our litmus test is non-trivial. 
Prior research has aimed to reverse engineer this mapping using RowHammer attacks~\cite{9152654, DRAMA, 10.1007/978-3-030-00470-5_3, googleprojectzero_2015, 197231}, other hardware fault injection~\cite{9152654, 8835222}, side-channel attacks~\cite{DRAMA}, reduced timing parameters~\cite{8615701, 10.1145/3084464}, thermal heaters~\cite{10.1145/2989081.2989114}, physical probing~\cite{DRAMA}, access-latency analysis~\cite{8494868}, or a combination of these techniques. 
Unfortunately, the available techniques are not suitable for our methodology due to their significant shortcomings. Our evaluation showed that none of the techniques are lightweight enough for our purpose, as they suffer from either inefficiency (taking hours to complete), invasiveness (posing a potential risk of damaging the chips), inconsistency (Rowhammer attacks are not always effective, and their success depends on the type of DIMMs used), need to be re-applied whenever the machine settings change, typically designed for single CPU architecture (mainly Intel), provide only approximate information, or often fail to output a deterministic DRAM address mapping.
In our context, we only need the mapping for a small set of physical addresses; we thus devised a simple software-only solution for our testing infrastructure, as we describe in \cref{sec:infrastructure}.


\section{Empirical Infrastructure}
\label{sec:infrastructure}

We now describe the novel testing infrastructure we have designed for empirically validating memory persistency models.

To gain insight into DDR bus activity and gain direct access to memory signals, we utilise a memory bus interposer, specifically the FS2800 DDR Detective \cite{FuturePlus}. 
This tool provides monitoring and close-proximity probing capabilities, enabling us to closely observe and analyse various signals on the DDR bus, such as activate, read, write, precharge, mode register write, mode register read, refresh, etc.
It also acts as a repeater, providing non-intrusive protocol decode of memory transactions and state analysis. Its logic analyser  serves as the analysis execution engine, commonly referred to as the "probe manager". The interposer probe can be easily set up by snapping it onto the motherboard of the target system and attaching the memory component to the top, as depicted schematically in \cref{fig:setup}.
Through this approach, it becomes possible to monitor traffic and process JEDEC\footnote{JEDEC (Joint Electron Device Engineering Council) is the engineering standardisation body that develops open standards and publications for the microelectronics industry, including DDR4.} DDR4 memory protocol requirements in real-time.
The bus analysis probe offers triggers for detecting state violations and user-selectable DDR commands or command sequences to enhance system analysis.
In addition, the testing infrastructure assumes that the Probe Manager host has the ability to establish a remote SSH tunnel with the target system.
The \DDRDetective{} presented four challenges for us:

\begin{figure*}[t!]
\captionsetup{width=1\textwidth}
\includegraphics[width=1\textwidth]{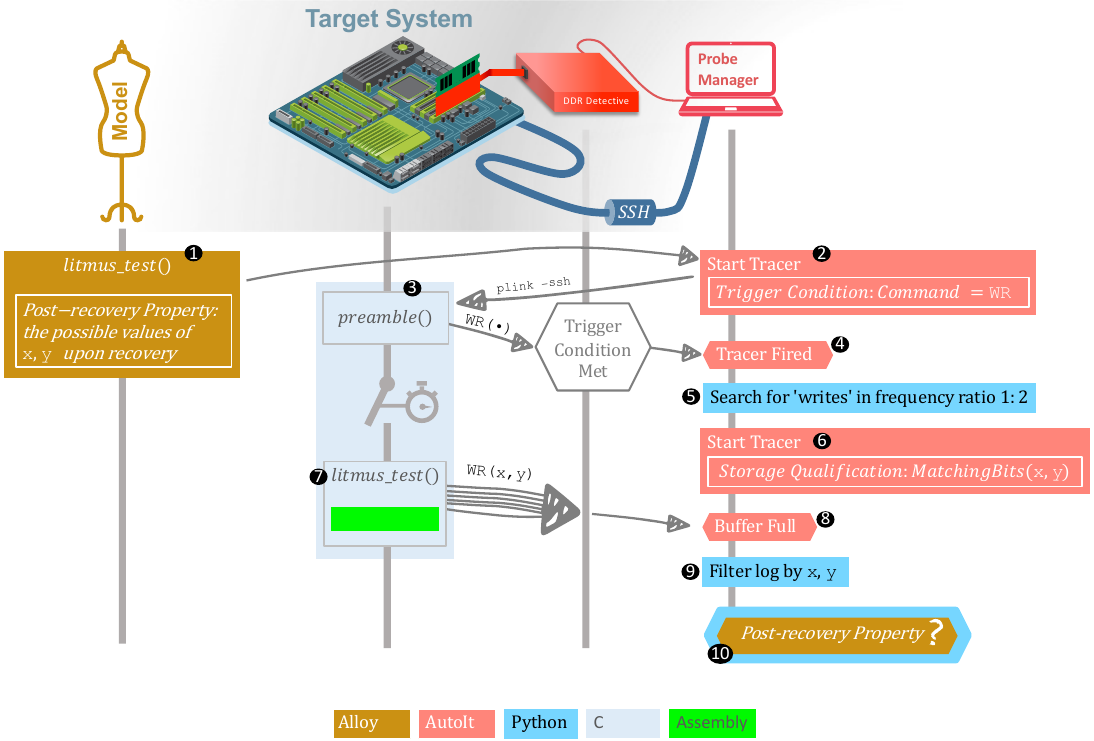}
  \caption{Our flow for validating persistency models; a colour nested inside another denotes an embedded program within outer code; a hexagon allows the flow to continue if the condition inside is true.}
  \label{fig:flow}
\end{figure*}

\begin{itemize}

\item
Our litmus tests refer to virtual memory addresses, but the addresses captured by \DDRDetective{} are actually geometric.

\item
The \DDRDetective{} can log the address of each memory write, but not the actual data (value) written. This means that we cannot infer the order of writes to the same location, only the order of writes to different locations.

\item
When the \DDRDetective{} is configured to record the address of \emph{every} write operation, it tends to exhaust its storage capacity quickly, depending on the size of the tracing log set. While it is possible to limit the logging of addresses by using a specific wildcard pattern, determining the appropriate pattern \emph{a priori} without prior knowledge is challenging.

\item
The \DDRDetective{} is controlled by a GUI application with no command-line support, making automation challenging.

\end{itemize}

We overcome these challenges as follows, where the numbers refer to the steps depicted in \cref{fig:flow}.

\begin{enumerate}

\item[\blacknum{1}]
The first step is to generate a litmus test. This takes the form of a small single- or multi-threaded program containing writes and reads on a fixed set of memory locations, possibly separated by persist or barrier instructions, together with a post-recovery condition constraining the allowed states of persistent memory. Raad \etal~\cite{RaadWV19,RaadWNV20} list several handwritten litmus tests for the Arm and Intel-x86 persistency models that can be used as a starting point. It is also possible to automatically generate litmus tests directly from the persistency model axioms, using Alloy-based techniques~\cite{lustig+17,RaadWNV20}. We, however, have considered only the simplest form of persistency litmus test -- a single thread containing two persist-separated writes to different locations (shown in \cref{alg:mthread}). This is because having observed violations of the persistency model on such a simple example, no additional insight can be gained by studying more elaborate examples.

\begin{litmustest}   

\caption[Caption for LOF]{A litmus test in ARMv8.2 that runs in a loop and is preceded by the corresponding preamble; the \texttt{asm} keyword declares inline assembler instructions that are embedded within the C code; $x_{pd}$ denotes the value of $x$ read from the persistence domain utilising physical probing.}\label{alg:mthread}

\hspace*{-.25cm} \includegraphics[width=\columnwidth]{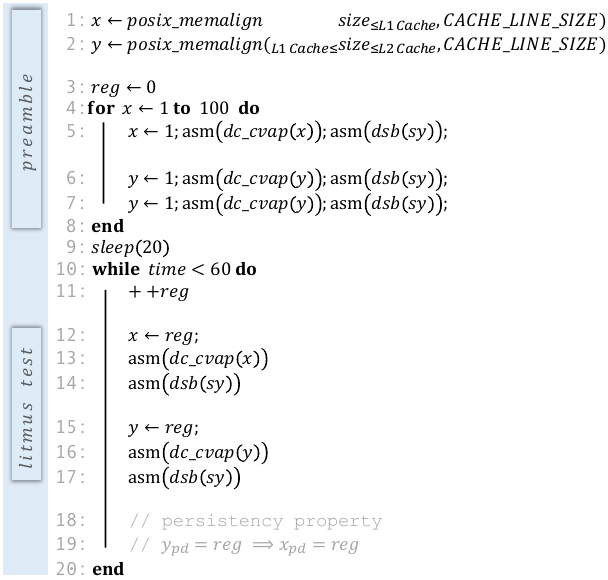}
\vspace*{.01cm}
 
\end{litmustest}

\item[\blacknum{2}]
  We now begin the process of determining which geometric addresses are in use. The logging process has been streamlined by instructing the \DDRDetective{} to selectively log only \singleq{writes} upon detecting the first one. Since the \DDRDetective{} logs \emph{all} \singleq{writes} occurring \emph{anywhere} in the system, the delay between the instruction to start logging when a write is observed and a write being observed is typically very short.

\item[\blacknum{3}]
  As soon as we have instructed the \DDRDetective{} to start logging, we launch the `preamble' of the litmus test. This allocates the locations that will be used in the test, and then performs a distinctive number of writes to each one. For instance, if there are two locations in use ({\it x} and {\it y}) then we perform 100 writes to {\it x} and 200 writes to {\it y}. After this, the litmus test pauses for about 20 seconds (the reason for this is explained in step \blacknum{4}).

\item[\blacknum{4}] Meanwhile, as soon the \DDRDetective{} observes a memory write somewhere in the system, it starts logging all memory writes. We configure the device in a mode where it logs a maximum of 8,000 writes, and then writes the log to a CSV file, which is analysed in step \blacknum{5}.\footnote{The log size, called the `trace memory depth' in FuturePlus terminology, is configurable; we found 8,000 writes ample for our experiments.} The reason for the 20 second pause in step \blacknum{3} is to allow plenty of time for the CSV log file to be produced and analysed.
  
\item[\blacknum{5}]
Our Python script then scans the log to identify the two memory locations that are written 100 and 200 times. This allows us to determine the relevant geometric addresses.

\item[\blacknum{6}]
Having now determined the relevant geometric addresses, we clear the \DDRDetective{} log. We then instruct the \DDRDetective{} to begin logging again, but only to log accesses whose address matches a particular pattern (there is no support for logging access to a specific set of addresses; only a single pattern can be provided). The pattern is the most specific one that matches all geometric addresses determined in step~\blacknum{5}. Consider a toy example where the geometric addresses are each four bits long, and we have three such addresses: $x= \texttt{0011}$, $y= \texttt{0101}$, and $z= \texttt{0001}$. In this case, we can use the pattern {\tt 0XX1} to identify a subset of addresses that match this pattern. The more bits the geometric addresses happen to have in common, the fewer wildcards (`{\tt X}' bits) in the pattern, the more the use of the pattern is able to reduce the amount of data logged, and the longer the litmus test can be run.
Our experiments indicate that the degree of overlap in the number of bits present in geometric addresses is heavily influenced by the memory size allocated by the litmus test to corresponding locations. On average, we observed 30 out of 32 bits\footnote{The DRAM component we used in our experiments features 18 row address bits, 10 column address bits, 2 bank address bits, and 2 bank group address bits.} in common when the memory size was small (in the range of a few bytes), while allocated memory sizes of a few MB resulted in an average of around 16 bits in common.

\item[\blacknum{7}]
The litmus test script is then reactivated. The main body of the litmus test performs as many iterations as possible within a fixed timeframe, which is 60 seconds in the \cref{alg:mthread}. We do not check whether the post-recovery condition of the litmus test holds on each iteration just yet, as this only becomes apparent once we inspect the log.

\item[\blacknum{8}]
The probe continuously captures data until the log reaches a predetermined size.

\item[\blacknum{9}]
We use another Python script to filter the log by removing all entries related to geometric addresses other than those we reverse-engineered in step~\blacknum{5}.

\item[\blacknum{10}]
We then inspect the log to check whether the post-recovery condition holds for each iteration of the litmus test. For example, in \cref{alg:mthread}, this is a matter of checking that the writes to {\it x} and {\it y} appear in strict alternation throughout the log. Two consecutive writes to {\it x} or {\it y} indicates either that the writes have been reordered or that a write has been added or deleted, and any of these indicate a violation of the persistency model.

\end{enumerate}

\paragraph{Automation}
The probe manager application of the \DDRDetective{} lacks command-line interface (CLI) support and relies solely on a Windows graphical user interface (GUI). This limitation poses a significant challenge for automation of experiments and remote access. To surmount this challenge, we resorted to leveraging AutoIt~\cite{autoit}, a scripting language designed to facilitate automation of the Windows GUI and general scripting tasks. AutoIt achieves this by simulating keystrokes, mouse movements as well as window and control manipulations, enabling effective automation of tasks that require interaction with the Windows GUI.

\subsection{Building confidence in our infrastructure}

We took several steps to build confidence that our empirical infrastructure does not produce false positives: that is, if it reports writes being persisted in a manner that violates the architecture's memory persistency model, then this is not simply an artefact of our setup. This is important because experience with litmus-testing of memory \emph{consistency} models indicates that violations can be very rare, perhaps observed only once every few hundred million runs~\cite[Figure~8]{AlglaveMSS10}.

\paragraph{Declaring memory locations as volatile}
We declared the locations used in the litmus test as `volatile', which instructs the C compiler not to reorder (or optimise away) any writes to these locations. (That said, when we tried omitting `volatile' we actually noticed no difference in the number of writes logged.)

\paragraph{Defending against start-up anomalies}
Were we to observe violations only at the very start or at the very end of a litmus test's execution, we might reasonably suspect that the violations were being triggered as a result of the litmus test starting or ending, rather than writes genuinely being persisted out-of-order. To defend against this, we noted not just the \emph{number} of anomalies in the log, but how these anomalies are distributed. \Cref{fig:distribution} shows a typical distribution of anomalies observed in our experiments; they can be seen to be reassuringly spread out over the 2000 writes.

\paragraph{Aligning blocks to cachelines}
During our initial testing we noted that the \DDRDetective{} often logged more writes to a location than our litmus test produced. We attributed this to the fact that multiple memory locations can share the same cacheline, and the whole cacheline is persisted to memory as a single unit. As such, we could be logging writes not just corresponding to the locations of interest, but also those corresponding to locations in the same cachelines as the locations of interest. To mitigate this, when allocating locations at the beginning of each litmus test, we made the allocation size a multiple of the cacheline size and aligned with a cacheline boundary, with the help of the {\tt posix\_memalign} function. By doing so, we could ensure that no instruction other than the one committed by the litmus test would reside in a cacheline of the allocated chunks.

Subject to these constraints, we also varied the size of the allocations, to see whether this had any impact on the persistency behaviour -- see \cref{sec:evaluation} for details and results of this experiment.

\paragraph{Cross-checking between two architectures}
To minimise the possibility of any interference caused by the DDR Detective affecting our results, we conducted cross-checks on two different architectures - Arm and Intel-x86 - using the same DDR Detective, settings, and litmus test on each run. Our analysis revealed that the Intel machine consistently exhibited negligible diversion of the number of persisted writes from the issued ones, whereas the Arm machine showed a significant and consistent diversion. 
We repeated the tests by modifying the DDR Detective settings and litmus tests, while keeping the DDR Detective device constant. 
However, the observed diversion in the Arm architecture remained significantly higher than in the Intel-x86 architecture. Therefore, we can conclude that if the DDR Detective had indeed introduced any interference, it would have been observable on the Intel architecture as well.


\section{Experimental Evaluation}\label{sec:evaluation}
We now present experiments that put the process depicted in  \cref{fig:flow} into practice, to ascertain 
\begin{inparaenum}[(1)]
\item the degree of adherence of the target machine to the specifications delineated in the vendors' manuals, with respect to the order in which the stores are propagated (flushed) to memory; and 
\item the veracity of the proposed persistency models.
\end{inparaenum}

Our experiments use the simple sequential program outlined in \cref{alg:sequential}, which builds upon \cref{alg:mthread} and offers a more comprehensive approach to automating the validation process. We selected this program because it is the smallest program that is potentially capable of exposing inconsistencies related to persistency semantics.
The program accesses two memory addresses, $x$ and $y$, in batches of $2{\small,}000$.
In between accessing these addresses, a persist instruction is executed, followed by a barrier. 
During this process, the \DDRDetective{} records the order in which data is transmitted through the memory bus, as illustrated in  \cref{fig:flow}.

\SetKwComment{tcp}{\textcolor{lightgray}{\footnotesize //} }{}%
\begin{litmustest}
\SetAlgoLined
\DontPrintSemicolon
\caption{A sequential program writing to two persistent memory locations $x, y$; the write on $x$ persists before that on $y$} \label{alg:sequential}

    \textcolor{lightgray}{\nl} \For{$i\gets 1$ \KwTo $2000$ }{
    \textcolor{lightgray}{\nl} $x \gets 1$\;
        \textcolor{lightgray}{\nl}   \For{$j\gets 1$ \KwTo $p$ }{
        \textcolor{lightgray}{\nl}      $persist(x)$\label{persistx}
             }
    \textcolor{lightgray}{\nl} $barrier$\;
    \textcolor{lightgray}{\nl} \textcolor{lightgray}{$sleep(\cdot)$} \tcp*[r]{\textcolor{lightgray}{\texttt{\small suspend execution}}} 
    
    \textcolor{lightgray}{\nl} $y \gets 1$\;
        \textcolor{lightgray}{\nl}   \For{$j\gets 1$ \KwTo $p$ }{
        \textcolor{lightgray}{\nl}      $persist(y)$\label{persisty}
            }
    \textcolor{lightgray}{\nl} $barrier$\;
    \textcolor{lightgray}{\nl} \textcolor{lightgray}{$sleep(\cdot)$ \tcp*[r]{\textcolor{lightgray}{\texttt{\small suspend execution}}}} 
    
    \textcolor{lightgray}{\nl}\vbox{\colorbox{CanaryYellow}{\vbox{$y=1 \implies x=1$ \tcp*[r]{\textcolor{lightgray}{\texttt{\small persistency property}}}}}}
    }
\vspace*{.2cm}
\end{litmustest}

\Cref{alg:sequential} is associated with a post-recovery condition (persistency property) that specifies the permissible values of $x$ and $y$ upon recovery. As mentioned earlier, this property cannot be checked programmatically. Therefore, we rely on the \DDRDetective{} to examine the order of the intercepted memory instructions that it logs.

\subsection{Metrics}

Our experimental assessment is based on two different metrics, namely, \emph{reorderings} and \emph{deviations}. 

\paragraph{Reorderings}
A reordering occurs when the expected order in which memory instructions in a program should be persisted is not maintained upon recovery. In other words, the sequence in which the instructions are observed as being committed to the persistence domain does not match the order that \emph{should} be observed according to the vendor's specification. Such reorderings may lead to incorrect behaviour, violating the intended logic of the program.

Verifying the post-recovery property can be reduced to checking the pattern of the ordering of writes that are propagated to the persistence domain.
Identifying violations of this pattern can present a significant challenge, especially for complex programs, as it can be arduous to isolate and comprehend the underlying factors that lead to reorderings. 
These factors are often intricately tied to the internal workings of the compiler or the processor itself, which can involve various known optimisations as well as other behaviours that are difficult to explain. 
Examples of such factors include instruction reorderings, instruction losses, and even the unexpected addition of instructions, as further demonstrated in the paper. 
Therefore, it can be a complex and challenging task to pinpoint the specific cause of an ordering violation.

Luckily, \cref{alg:sequential} is expected to show a simple alternating pattern between the writes to $x$ and $y$. To simplify the definition of ``reorderings'' for this test, we count the number of times that two consecutive writes occur to the same location, as observed by the \DDRDetective{}.

\paragraph{Deviation}
To quantify the discrepancy between the number of writes actually persisted, $P(X)$, to persistent memory locations $X = x_1, x_2, \dots, x_n$, as observed and logged by the interposer, and the number of writes issued by the test program, $I(X)$, we utilise the metric $\Delta_{P(X)}^{I(X)}$, denoted as $\Delta X$ for simplicity. 
This metric is defined as the following (percent) normalised unsigned deviation:

$$\Delta_{P(X)}^{I(X)} = \sum_{i=1}^{n} \frac{|P(x_i) - I(x_i)|}{I(x_i)} \times 100 \% $$
Here, the sum goes over each memory location $x_i$ and computes the absolute difference between the number of writes recorded by the interposer, $P(x_i)$, and the number of writes issued by the test program, $I(x_i)$. The deviation is normalised by the number of writes issued by the test program.
A value of 0 indicates perfect agreement between the issued writes and the persisted ones, while higher values indicate increasing levels of disagreement. 
The term $\frac{P(x_i) - I(x_i)}{I(x_i)} $ denotes the signed  deviation of the persists to the individual memory location $x_i$ from the number of writes issued to that location, denoted by $\Delta x_i$.

For instance, suppose we run \cref{alg:sequential} 2,000 times (hence the issued writes to the persistence domain are 2,000 for both $x$ and $y$), and the \DDRDetective{} detects 1,950 writes to memory location $x$ and 2,050 writes to memory location $y$. In this scenario, the deviation can be computed as follows:
\[
\Delta xy = \Delta_{\{1950,2050\}}^{\{2000,2000\}} = \frac{|1950 - 2000| + |2050 - 2000|}{2000}  = 5\%
\]
The amount of signed deviation for $x$ is $\Delta x = -2.5\%$, which implies that the number of observations for $x$ is lower than expected. Conversely, the signed deviation for $y$ is $\Delta y = +2.5\%$, suggesting that the number of observations for $y$ is higher than expected.

\subsection{Machines under test}
\label{sec:machines}

The experiments on the Intel-x86 architecture were conducted on an Intel\textsuperscript{\sffamily\textregistered} Xeon\textsuperscript{\sffamily\textregistered} CPU E5-1630 v3, running at 3.70GHz (Quad-Core) with 128 KiB L1 cache, 1 MiB L2 cache and 10 MiB L3 cache.

The experiments for the ARM architecture were conducted on an Ampere\textsuperscript{\sffamily\textregistered} Altra\textsuperscript{\sffamily\texttrademark} M96-30 System-On-Chip (SOC) featuring 96 Arm v8.2+ Cores, each with a 64 KB L1 I-cache, 64 KB L1 D-cache, and a 1 MB L2 cache. 

\subsection{Key Observations}\label{sec:keyobservations}

\subsubsection{Intel-x86}

According to Intel~\cite{wpq} \cite[p.~3]{Rudoff}, writes destined for persistent memory become persistent once they enter the write-pending queue (WPQ) in the memory controller, which is battery-backed. Therefore we knew that if we observed reorderings using the \DDRDetective{} on our Intel-x86 machine, we could not necessarily attribute these to persistency violations. Nevertheless, we decided it would be interesting to see whether we could in fact detect such reorderings using our setup.

We ran \cref{alg:sequential} 10 times on our Intel machine, and on average, we observed deviations of only 0.04\% between $x$ and $y$, with barely one instance of instruction reordering.

We discussed our findings with a Senior Software Engineer at Intel, and they confirmed that:
\begin{shadequote}{}%
The persistence domain includes the write pending queues in the memory controller. A store is considered persistent once it reaches that queue. At that point, there’s no reason to maintain any ordering information since power loss will flush the queue completely (and in any order).
\end{shadequote}

We believe that, in their present form, there is no reliable way for the persistency behaviour of Intel machines to be independently validated: persistent programmers have no choice but to simply trust that these machines have been implemented in accordance with the architectural specification. This is an important shortcoming that our work is, to our knowledge, the first to bring to light.

\subsubsection{Arm}

In contrast to Intel-x86, Arm machines are not expected to have a battery-backed memory controller. Therefore, given that the \cvap{} instruction is designed for pushing data out to the Point of Persistence (PoP), we were confident that our approach would work if we could find an Arm machine that implements \cvap.

We contacted a persistency expert at Arm, who advised us to purchase the Ampere\textsuperscript{\sffamily\textregistered} Altra\textsuperscript{\sffamily\texttrademark}, writing:
\begin{shadequote}{}%
There are Arm-based systems out there such as the Ampere  Altra, which features Armv8.2 compliant cores, where \cvap{} support is mandatory. On the other hand, we've not tested how that system does persist operations.
\end{shadequote}

Based on this advice we procured the Arm machine described in \cref{sec:machines} and applied our experimental setup to it.
Again, we ran \cref{alg:sequential} 10 times, and we were therefore surprised to observe a substantial number of reorderings and deviations. On average, we observed deviations of 2.5\% between $x$ and $y$, and 5.78\% reorderings. 

We contacted Ampere to discuss our findings with them. They advised us that -- contrary to our expectations -- the Altra machine does \emph{not} have any support for persistent memory. They wrote:
\begin{shadequote}{}%
We currently do not support building fine grained persistent memory platform solutions with the Ampere\textsuperscript{\sffamily\textregistered} Altra\textsuperscript{\sffamily\texttrademark} SOCs. [\ldots] That given, the Ampere\textsuperscript{\sffamily\textregistered} Altra\textsuperscript{\sffamily\texttrademark} SOC does not have a Point of Persistence.
\end{shadequote}
They pointed us to a loophole in the Arm 8.2 architecture that allows the \cvap{} to simply behave like a \cvac{} if the system does not define a Point of Persistency.
\begin{shadequote}{}%
When {\tt FEAT\_DPB} is implemented, meaning the \cvap{} instruction is implemented, if the memory system does not support the Point of Persistence, a data cache clean to the PoP, \cvap{}, behaves as a data cache clean to the PoC, \cvac.~\cite[p.~5057]{ARMmanual}
\end{shadequote}

Our understanding is that a programmer can query the {\tt FEAT\_DPB} to confirm that a system supports the \cvap{} instruction. However, there is no way for a programmer to query where a system's Point of Persistence is, or indeed, whether it defines one at all. Therefore, there is no way for a programmer to know whether the \cvap{} is implemented `meaningfully' or not, without talking to the system designer directly.

To enhance our understanding of the behaviour of \cvap{}, we further analysed the distribution of reorderings for each memory location in the litmus test, which spanned 2{\small,}000 executions. Our objective was to identify any discernible patterns or trends in the occurrence of reorderings that might provide further insight into the behaviour of the \cvap.
A blue horizontal line in the distribution plot in \cref{fig:distribution} indicates that a write to memory location $x$ of \cref{alg:sequential} was issued by the program but not logged in the \DDRDetective{} tracer. We identify this by observing two consecutive writes to location $y$ in the same iteration of the litmus test.
Similarly, the orange line represents the case where a write to location $y$ was expected but not logged in the \DDRDetective{} tracer.
As shown in \cref{fig:distribution}, we observed reorderings across the entire spectrum of 2{\small,}000 executions. However, the majority of these reorderings were concentrated consistently in the first half of the executions, with a higher density in this portion of the dataset. 

One of the most concerning observations we made on the ARM machine was that we frequently observed more writes being intercepted by the \DDRDetective{} than were issued by the litmus test. We were unable to find a satisfactory explanation for this phenomenon, while fewer propagated writes could potentially be justified by compiler optimisations, for example.

\subsection{Curiosity-driven experiments}

The negative findings of \cref{sec:keyobservations} meant that it would not be meaningful to pursue a full-blown persistency litmus testing campaign, given that reorderings were observed for even the simplest of litmus tests.

We decided to instead focus on using or DDR Detective-based setup to perform some curiosity-driven experiments to assess properties of our Arm machine. These experiments demonstrate that our setup can be used to answer nuanced questions related to memory traffic, providing confidence that when Arm machines that \emph{do} support persistent memory are produced in due course, a setup like ours is likely to be useful in their empirical validation.

\paragraph{Does DC CVAP degenerate to DC CVAC?}
Our understanding is that the loophole in the Arm 8.2 architecture for systems that do not have a point of persistence is that the \cvap{} instruction should simply behave like \cvac. We set about using our DDR Detective-based setup to assess whether there is any noticeable difference in observed reorderings when \cvac{} is used in place of \cvap{} in litmus tests running on the Arm machine.

\begin{figure}
    \centering
         \begin{subfigure}[b]{0.49\columnwidth}
         \centering         \includegraphics[width=\columnwidth]{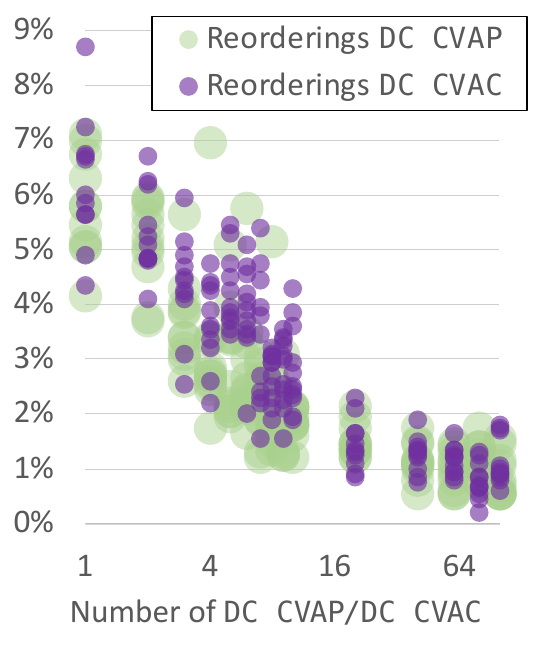}
         \caption{}         \label{fig:reord_cvap_cvac}
     \end{subfigure}
     \hfill
     \begin{subfigure}[b]{0.49\columnwidth}
         \centering
         \includegraphics[width=\columnwidth]{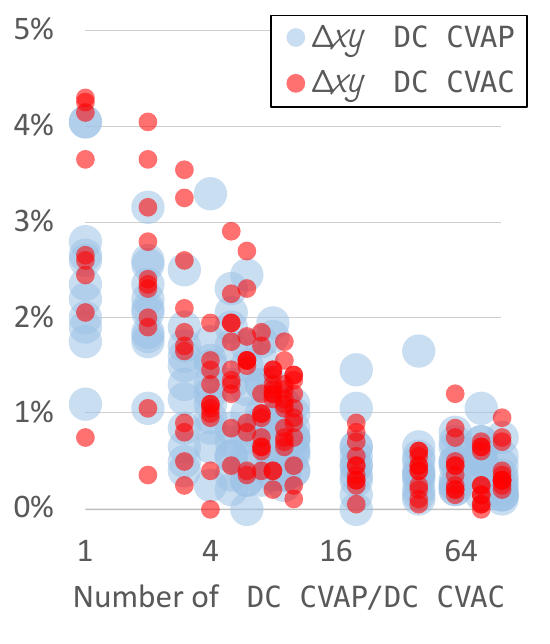}
         \caption{}
         \label{fig:delta_cvap_cvac}
     \end{subfigure}
        \caption{\emph{Reorderings} (left) and \emph{deviations} (right) as the number of \cvap/\cvac{} varies in \cref{alg:sequential} -- each data point on the graph represents the results of a 2K-iteration run of the test on the Arm machine}
        \label{fig:cvap_vs_cvac}
\end{figure}

We performed a series of experiments where we incrementally added successive \cvap{}s in \cref{alg:sequential}, \cref{persistx,persisty}, before repeating the process for \cvac{}. 
The outcomes of these experiments are presented in \cref{fig:cvap_vs_cvac}.
We performed a Mann-Whitney U test with a significance level of 0.05 to determine the statistical significance of the difference in the occurences of \emph{reorderings} and \emph{deviations} between the populations using an equal number of \cvap{} and \cvac{} instructions. The test results exhibited mixed behaviour. Specifically, for \emph{reorderings}, the two population samples were statistically equal when using 1, 2, 20, 40, and 100 \cvap{}/\cvac{} instructions, but were statistically unequal in other cases (using 3-10, 60, and 80 instructions).
In terms of the \emph{deviations} metric, the Mann-Whitney U test revealed a higher degree of consistency in statistical equivalence between the samples, with only one instance (using 9 instructions) showing non-equivalence.
Although we observed a higher degree of non-equivalence for the reorderings metric compared to deviations, our statistical analysis did not yield sufficient evidence to reject the assumption that the behaviour of the \cvap{} instruction defaults to that of \cvac.

\begin{figure}
    \centering
         \begin{subfigure}[b]{0.50\columnwidth}
         \centering         \includegraphics[width=\columnwidth]{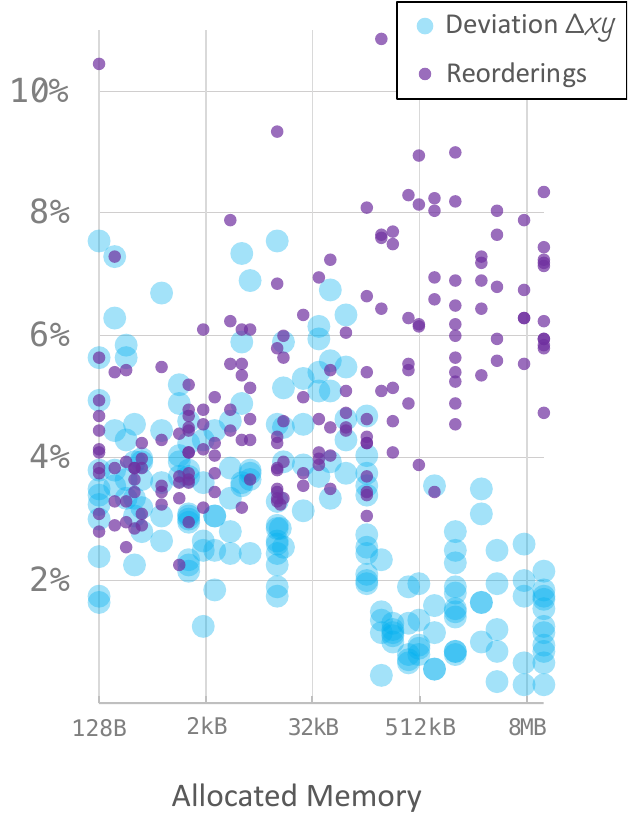}
         \caption{}         \label{fig:increaseMEM_devANDreord}
     \end{subfigure}
     \hfill
     \begin{subfigure}[b]{0.48\columnwidth}
         \centering
         \includegraphics[width=\columnwidth]{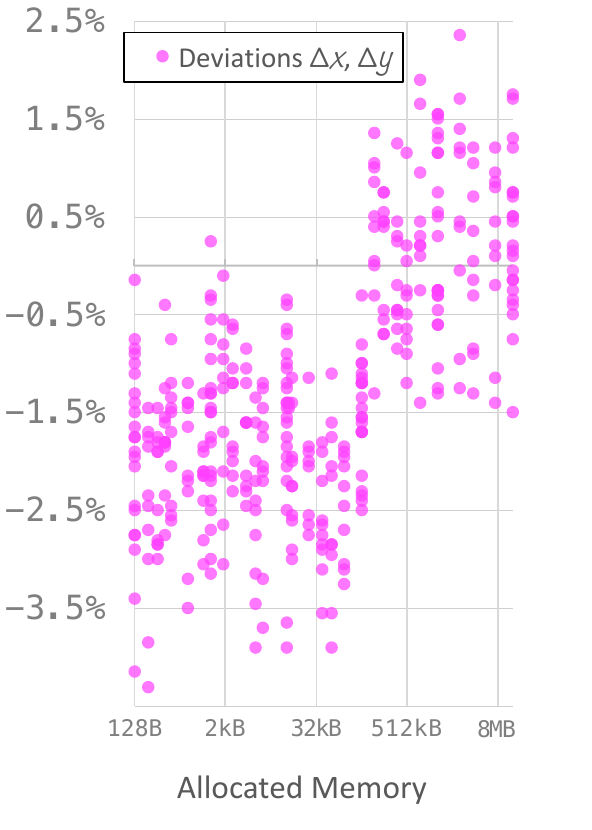}
         \caption{}
         \label{fig:increaseMEM_absdev}
     \end{subfigure}
        \caption{\emph{Deviations/Reorderings} (left) and \emph{Signed Deviations} (right) as the size of the allocated memory varies uniformly across all memory locations; each data point on the graph represents the results of a 2K-iteration run of the litmus test on Arm machine}
        \label{fig:increaseMEM}
\end{figure}

\paragraph{Does the distance between the memory locations affect reorderings?}
We also noted a correlation between the size of memory allocation for a given location and the frequency of reorderings, as well as the deviation observed at that location. Specifically, as shown in \cref{fig:increaseMEM}, we found that memory locations with larger allocations tended to experience a higher number of reorderings than those with smaller allocations. Conversely, we observed that the locations with smaller allocations tended to exhibit higher deviations, while the locations with larger allocations experienced lower deviations.
To further investigate this trend, we plotted the signed deviations of the intercepted persists (both for $x$ and $y$) from the number of issued writes in \cref{fig:increaseMEM_absdev}. 
As the allocated memory size for a specific location grew, we noticed a rise in the number of intercepted persists, which, although initially less than the number of writes issued by the litmus test (plot marks below the horizontal axis), eventually exceeded it.
Notably, the graphs exhibit a distinct shift after 64 KB, coinciding with the L1 cache size per core.

\begin{figure}
    \centering

             \begin{subfigure}[b]{0.72\columnwidth}
         \centering         \includegraphics[width=\columnwidth]{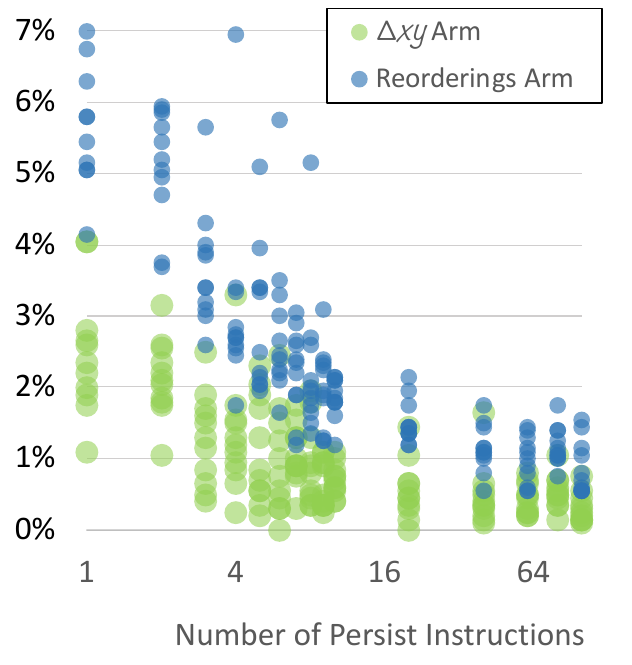}
         \caption{}         \label{fig:invalidations_graph}
     \end{subfigure}
      \hfill
     \begin{subfigure}[b]{0.27\columnwidth}
         \centering
         \includegraphics[width=\columnwidth]{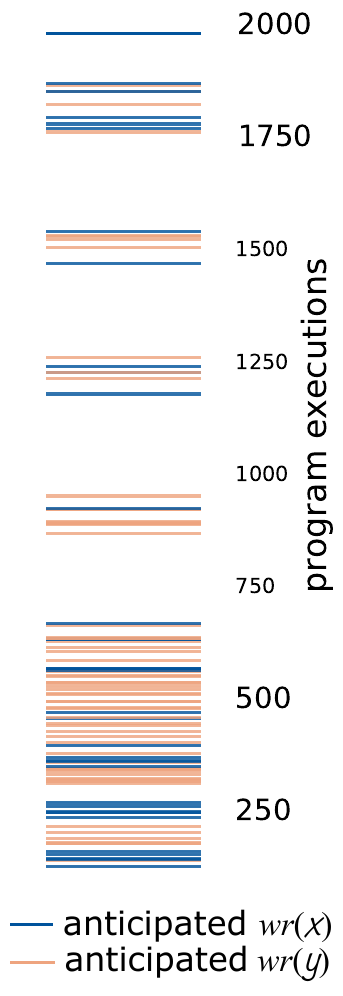}
         \caption{}
         \label{fig:distribution}
     \end{subfigure}

            \caption{\emph{Reorderings/deviations} as the number of persists varies in \cref{alg:sequential} -- each mark on the graph represents a 2K-iteration run of the test (left);
            and
            a random \emph{distribution} of the anomalies of a single (2K-iteration) run of \cref{alg:sequential} with a single persist (right). }
        \label{fig:graphs}
\end{figure}

\paragraph{Do repeated persists inhibit reorderings?}
To gain insight into the  underlying semantics of \cvap, as observed by our validation framework, we conducted experiments in which we gradually increased the number of consecutive \cvaps{} from 1 to 100 within \cref{alg:sequential}.
The results, shown in \cref{fig:invalidations_graph}, demonstrate that an increase in the number of \emph{persist} (i.e., \cvaps) is accompanied by an exponential decrease in both reorderings and deviation. These results imply that including multiple \emph{persist} instructions can enhance the alignment of the observed semantics of \cvap{} with the specifications.

\begin{figure}
    \centering
         \begin{subfigure}[b]{0.465\columnwidth}
         \centering         \includegraphics[width=\columnwidth]{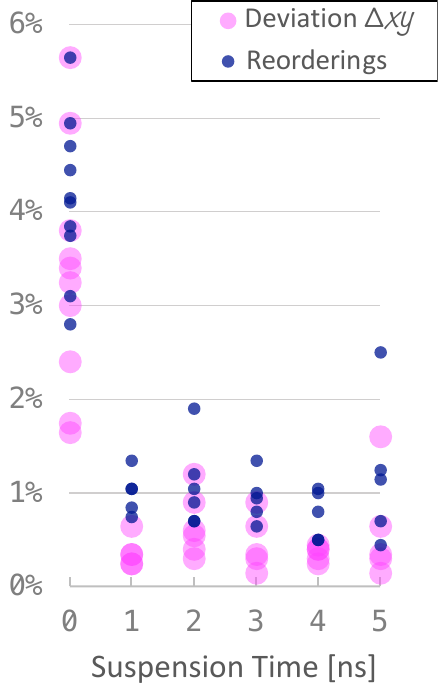}
         \caption{}         \label{fig:susp_reord}
     \end{subfigure}
     \hfill
     \begin{subfigure}[b]{0.525\columnwidth}
         \centering
         \includegraphics[width=\columnwidth]{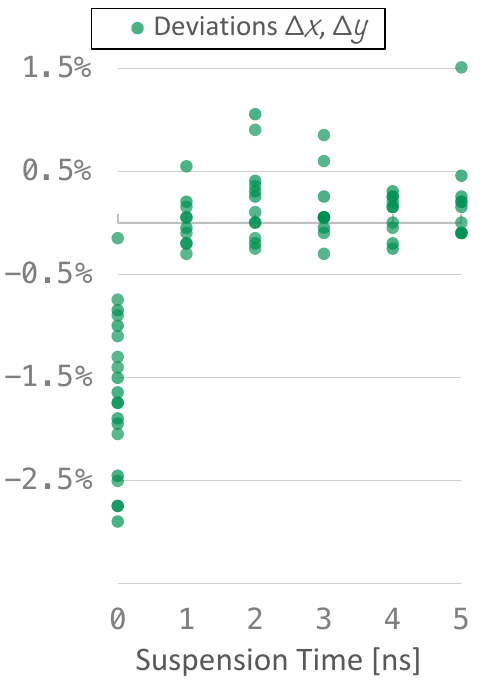}
         \caption{}
         \label{fig:susp_signed}
     \end{subfigure}
        \caption{\emph{Deviations/Reorderings} (left) and \emph{Signed Deviations} (right) observed during the execution of  \cref{alg:sequential} with intermittent suspension after each issued write.}
        \label{fig:increaseSUSP}
\end{figure}

\paragraph{Does suspending the processor between writes inhibit reorderings?}
We hypothesised that slowing down the program's execution by introducing delays between memory accesses (suspending the thread from execution for a short time) would reduce the occurrence of reorderings and deviations. 
To test this hypothesis, we repeated the experiment using \cref{alg:sequential} and added delays after each \emph{write-persist-barrier} sequence. 
Both $x$ and $y$ locations were allocated 100 bytes.
As shown in \cref{fig:susp_reord}, the delays significantly reduced the number of reorderings and deviations when compared to running the program without any delays. Moreover, this reduction was consistent across all delay times, ranging from 1 nanosecond upwards, without any noteworthy fluctuations as the delay time increased. These results suggest that the presence of a delay, rather than its specific duration, is the critical factor for this shift.
According to \cref{fig:susp_signed}, the signed deviations of the persist operations to both memory locations in the litmus test around the expected writes (2,000) decrease significantly with the introduction of suspensions. Both graphs in \cref{fig:increaseSUSP} clearly support that the behaviour observed with delayed writes is more in line with the vendor specifications for \cvap, although it is not an exact match.
Out of curiosity, we also attempted adding other random instructions between the memory accesses instead of using suspensions. However, we found that the anomalies remained unchanged compared to setting zero suspension time.

\paragraph{Does the value written make a difference?}
As part of our experiments, we tested two approaches for assigning values to memory locations: fixed values (e.g. $x:=1$) on each iteration, and varying values (such as an incrementing counter) on each iteration. Our objective was to identify potential compiler optimisations that might impact the reorderings and variability of persists. However, we found no noticeable difference between the two methods.

\medskip
In summary, this range of curiosity-driven experiments demonstrate the flexibility of our DDR Detective-based setup in allowing various questions related to the order in which writes are committed to memory to be assessed. Although Arm systems with support for persistency turn out not to be available at present, our approach sets the stage for empirical validation of their persistency guarantees once they do become available.
\section{Related Work}
\label{sec:related}

\paragraph{Validating memory consistency models}
To the best of our knowledge, our work marks the first attempt to empirically validate a memory \emph{persistency} model. We build upon more than a decade of work for empirically validating memory \emph{consistency} models for various architectures, including x86~\cite{litmus, SewellSONM10}, IBM Power~\cite{SarkarSAMW11}, and Arm~\cite{AlglaveDGHM21} multiprocessors, as well as GPUs~\cite{AlglaveBDGKPSW15} and hybrid CPU/FPGA chips~\cite{IorgaDSW21}.
Recent years have seen these models (and their validation testbenches) being extended to handle more architectural features, such as virtual memory~\cite{SimnerAPPGS22} and non-temporal memory accesses~\cite{RaadMV22}.
All of these consistency models can be validated purely in software, by having several threads perform memory accesses and noting which values are observed. That approach does not extend to the validation of \emph{persistency} models because a program has no way to tell whether it observes a value directly from persistent memory or simply from a volatile cache.

\paragraph{Deep persistency in the Arm architecture}
Recent proposals from Arm involve extending the persistence domain to include battery-backed memory controllers or caches \cite{AlshboulRWTS21}. In such systems, the point of persistency (PoP) moves to between the caches and the memory controller \cite[p.~8]{wang22}, which brings it into line with Intel-x86. Version 8.5 of the Arm architecture defines an additional point of \emph{deep} persistence (PoDP) for such systems, which lies between the memory controller and the persistent memory. The idea is that when data goes beyond the PoDP, it is  ``more strongly'' guaranteed to be kept in the event of a power failure~\cite[p.~9]{deep_persistence}. The proposed $\mathit{dc\_cvadp}$ instruction propagates writes to the PoDP (in contrast to $\mathit{dc\_cvap}$ which only guarantees propagation to the PoP). 
This means that our approach could be applied to an Arm system that contains battery-backed components, providing that it supports the 
 $\mathit{dc\_cvadp}$ instruction.

\paragraph{Testing persistent programs}
Complementing our work on persistency model validation, is a line of work on testing the persistent programs that run on them. 
For instance, PMTest~\cite{pmtest} is a framework for finding bugs in persistent programs. It works by having the user annotate their program with assertions of the form ``this write persists before that write'' or ``the write to $X$ has persisted by this program point'', and then seeking to invalidate these assertions via dynamic analysis. Pmemcheck~\cite{pmemcheck} and PMAT~\cite{pmat} work on a similar basis. XFDetector~\cite{xfdetector} goes further by searching for unintended interactions between instructions executed before and after a crash. PMFuzz~\cite{pmfuzz} is a test-case generator for persistent programs that seeks to maximise path-coverage, particularly paths that include persistency-related instructions such as write-backs. All of these tools use the vendor's architectural specification as the ground truth for determining whether a program has a bug, and hence they all benefit from our efforts to make the architectural specifications more trustworthy.

\section{Conclusions and Future Work}\label{sec:conclusion}

We have introduced a litmus-testing campaign with the aim of empirically validating the persistency guarantees of Intel-x86 and Arm machines. Since traditional validation methods used for memory consistency models do not directly apply to this domain, our approach played a key role in demonstrating the feasibility of validating memory persistency models. Through our experimental setup, we have provided empirical evidence that suggests either a lack of adherence to the specifications outlined in the vendors' manuals, or the absence of a reliable means to independently validate the persistency behaviour. Our findings contribute to a better understanding of the persistency behaviour of modern architectures and highlight the importance of testing the actual behaviour of memory persistency subsystems in contrast to relying solely on the vendors' documentation.

We believe that there is significant potential for applying intelligent validation techniques in the field of memory persistency. In our future work, we plan to explore an alternative approach for testing the persistency behavior of Arm processors, which involves using systems-on-chip, such as Xilinx's Zynq Ultrascale+. These chips combine a multicore Arm processor with a region of programmable logic (FPGA). 
%

\begin{figure}
    \centering
    \includegraphics[width=\columnwidth]{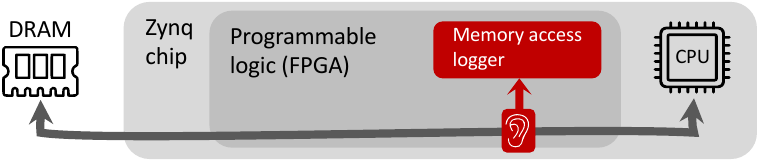}
    \caption{An alternative experimental setup suitable for testing embedded Arm processors}
    \label{fig:alternative_setup}
\end{figure}

As shown in \cref{fig:alternative_setup}, it may be possible to route memory accesses from the Arm processor via the programmable logic, so that those accesses can be recorded as they pass by a logger implemented in hardware. Indeed, the LiME framework of Jain et al.~\cite{JainLG18} uses a similar setup, except that the programmable logic emulates the DRAM rather than just transparently passing data. However, to our knowledge, there are no systems-on-chip currently available that have support for the Arm v8.2 architecture, which is the first to define the primitives required for persistent programming.


\bibliographystyle{IEEEtranS}
\bibliography{references}

\begin{thebibliography}{10}
\providecommand{\url}[1]{#1}
\csname url@samestyle\endcsname
\providecommand{\newblock}{\relax}
\providecommand{\bibinfo}[2]{#2}
\providecommand{\BIBentrySTDinterwordspacing}{\spaceskip=0pt\relax}
\providecommand{\BIBentryALTinterwordstretchfactor}{4}
\providecommand{\BIBentryALTinterwordspacing}{\spaceskip=\fontdimen2\font plus
\BIBentryALTinterwordstretchfactor\fontdimen3\font minus \fontdimen4\font\relax}
\providecommand{\BIBforeignlanguage}[2]{{%
\expandafter\ifx\csname l@#1\endcsname\relax
\typeout{** WARNING: IEEEtranS.bst: No hyphenation pattern has been}%
\typeout{** loaded for the language `#1'. Using the pattern for}%
\typeout{** the default language instead.}%
\else
\language=\csname l@#1\endcsname
\fi
#2}}
\providecommand{\BIBdecl}{\relax}
\BIBdecl

\bibitem{AlglaveBDGKPSW15}
\BIBentryALTinterwordspacing
J.~Alglave, M.~Batty, A.~F. Donaldson, G.~Gopalakrishnan, J.~Ketema, D.~Poetzl, T.~Sorensen, and J.~Wickerson, ``{GPU} concurrency: Weak behaviours and programming assumptions,'' in \emph{Proceedings of the Twentieth International Conference on Architectural Support for Programming Languages and Operating Systems, {ASPLOS} 2015, Istanbul, Turkey, March 14-18, 2015}, {\"{O}}.~{\"{O}}zturk, K.~Ebcioglu, and S.~Dwarkadas, Eds.\hskip 1em plus 0.5em minus 0.4em\relax {ACM}, 2015, pp. 577--591.
\BIBentrySTDinterwordspacing

\bibitem{AlglaveDGHM21}
\BIBentryALTinterwordspacing
J.~Alglave, W.~Deacon, R.~Grisenthwaite, A.~Hacquard, and L.~Maranget, ``Armed cats: Formal concurrency modelling at arm,'' \emph{{ACM} Trans. Program. Lang. Syst.}, vol.~43, no.~2, pp. 8:1--8:54, 2021.
\BIBentrySTDinterwordspacing

\bibitem{AlglaveMSS10}
\BIBentryALTinterwordspacing
J.~Alglave, L.~Maranget, S.~Sarkar, and P.~Sewell, ``Fences in weak memory models,'' in \emph{Computer Aided Verification, 22nd International Conference, {CAV} 2010, Edinburgh, UK, July 15-19, 2010. Proceedings}, ser. Lecture Notes in Computer Science, T.~Touili, B.~Cook, and P.~B. Jackson, Eds., vol. 6174.\hskip 1em plus 0.5em minus 0.4em\relax Springer, 2010, pp. 258--272.
\BIBentrySTDinterwordspacing

\bibitem{litmus}
\BIBentryALTinterwordspacing
J.~Alglave, L.~Maranget, S.~Sarkar, and P.~Sewell, ``Litmus: Running tests against hardware,'' in \emph{Tools and Algorithms for the Construction and Analysis of Systems - 17th International Conference, {TACAS} 2011, Held as Part of the Joint European Conferences on Theory and Practice of Software, {ETAPS} 2011, Saarbr{\"{u}}cken, Germany, March 26-April 3, 2011. Proceedings}, ser. Lecture Notes in Computer Science, P.~A. Abdulla and K.~R.~M. Leino, Eds., vol. 6605.\hskip 1em plus 0.5em minus 0.4em\relax Springer, 2011, pp. 41--44.
\BIBentrySTDinterwordspacing

\bibitem{AlshboulRWTS21}
\BIBentryALTinterwordspacing
M.~A. Alshboul, P.~Ramrakhyani, W.~Wang, J.~Tuck, and Y.~Solihin, ``{BBB:} simplifying persistent programming using battery-backed buffers,'' in \emph{{IEEE} International Symposium on High-Performance Computer Architecture, {HPCA} 2021, Seoul, South Korea, February 27 - March 3, 2021}.\hskip 1em plus 0.5em minus 0.4em\relax {IEEE}, 2021, pp. 111--124.
\BIBentrySTDinterwordspacing

\bibitem{ARMmanual}
\BIBentryALTinterwordspacing
{armDeveloper}, ``{Arm Architecture Reference Manual for A-profile architecture},'' 2022, \url{https://developer.arm.com/documentation/ddi0487/latest}.
\BIBentrySTDinterwordspacing

\bibitem{autoit}
\BIBentryALTinterwordspacing
{AutoIt}, ``{AutoIt Downloads},'' 2022, \url{https://www.autoitscript.com/site/autoit/downloads/}.
\BIBentrySTDinterwordspacing

\bibitem{8494868}
A.~Barenghi, L.~Breveglieri, N.~Izzo, and G.~Pelosi, ``Software-only reverse engineering of physical dram mappings for rowhammer attacks,'' in \emph{2018 IEEE 3rd International Verification and Security Workshop (IVSW)}, 2018, pp. 19--24.

\bibitem{arm_persistency_announcement}
\BIBentryALTinterwordspacing
D.~Brash, ``Armv8-a architecture evolution,'' 2016.
\BIBentrySTDinterwordspacing

\bibitem{kuco}
\BIBentryALTinterwordspacing
Y.~Chen, Y.~Lu, B.~Zhu, A.~C. Arpaci{-}Dusseau, R.~H. Arpaci{-}Dusseau, and J.~Shu, ``Scalable persistent memory file system with kernel-userspace collaboration,'' in \emph{19th {USENIX} Conference on File and Storage Technologies, {FAST} 2021, February 23-25, 2021}, M.~K. Aguilera and G.~Yadgar, Eds.\hskip 1em plus 0.5em minus 0.4em\relax {USENIX} Association, 2021, pp. 81--95.
\BIBentrySTDinterwordspacing

\bibitem{ChoLRK21}
\BIBentryALTinterwordspacing
K.~Cho, S.-H. Lee, A.~Raad, and J.~Kang, ``Revamping hardware persistency models: View-based and axiomatic persistency models for intel-x86 and armv8,'' in \emph{Proceedings of the 42nd ACM SIGPLAN International Conference on Programming Language Design and Implementation}, ser. PLDI 2021.\hskip 1em plus 0.5em minus 0.4em\relax New York, NY, USA: Association for Computing Machinery, 2021, p. 16â€“31.
\BIBentrySTDinterwordspacing

\bibitem{techinsightsblog}
\BIBentryALTinterwordspacing
J.~Choe, ``Review and things to know: Flash memory summit 2022,'' \emph{TechInsights}, August 2022.
\BIBentrySTDinterwordspacing

\bibitem{9152654}
L.~Cojocar, J.~Kim, M.~Patel, L.~Tsai, S.~Saroiu, A.~Wolman, and O.~Mutlu, ``Are we susceptible to rowhammer? an end-to-end methodology for cloud providers,'' in \emph{2020 IEEE Symposium on Security and Privacy (SP)}, 2020, pp. 712--728.

\bibitem{8835222}
L.~Cojocar, K.~Razavi, C.~Giuffrida, and H.~Bos, ``Exploiting correcting codes: On the effectiveness of ecc memory against rowhammer attacks,'' in \emph{2019 IEEE Symposium on Security and Privacy (SP)}, 2019, pp. 55--71.

\bibitem{FuturePlus}
\BIBentryALTinterwordspacing
{FuturePlus Systems}, ``{DDR Detective FS2800},'' 2017, \url{https://www.futureplus.com/ddr-detective/}.
\BIBentrySTDinterwordspacing

\bibitem{googleprojectzero_2015}
{Google Project Zero}, ``Exploiting the dram rowhammer bug to gain kernel privileges,'' 2015, \url{https://googleprojectzero.blogspot.com/2015/03/exploiting-dram-rowhammer-bug-to-gain.html}.

\bibitem{wpq}
\BIBentryALTinterwordspacing
{Intel}, ``{Deprecating the PCOMMIT Instruction},'' 2016, \url{https://www.intel.com/content/www/us/en/developer/articles/technical/deprecate-pcommit-instruction.html}.
\BIBentrySTDinterwordspacing

\bibitem{IorgaDSW21}
\BIBentryALTinterwordspacing
D.~Iorga, A.~F. Donaldson, T.~Sorensen, and J.~Wickerson, ``The semantics of shared memory in intel {CPU/FPGA} systems,'' \emph{Proc. {ACM} Program. Lang.}, vol.~5, no. {OOPSLA}, pp. 1--28, 2021.
\BIBentrySTDinterwordspacing

\bibitem{JainLG18}
\BIBentryALTinterwordspacing
A.~K. Jain, G.~S. Lloyd, and M.~B. Gokhale, ``Microscope on memory: Mpsoc-enabled computer memory system assessments,'' in \emph{26th {IEEE} Annual International Symposium on Field-Programmable Custom Computing Machines, {FCCM} 2018, Boulder, CO, USA, April 29 - May 1, 2018}.\hskip 1em plus 0.5em minus 0.4em\relax {IEEE} Computer Society, 2018, pp. 173--180.
\BIBentrySTDinterwordspacing

\bibitem{pmat}
L.~Jenkins and M.~L. Scott, ``Persistent memory analysis tool ({PMAT}),'' in \emph{11th Annual Non-Volatile Memories Workshop}, 2020, \url{https://louisjenkinscs.github.io/publications/PMAT_EA.pdf}.

\bibitem{10.1145/2989081.2989114}
\BIBentryALTinterwordspacing
M.~Jung, C.~C. Rheinl\"{a}nder, C.~Weis, and N.~Wehn, ``Reverse engineering of drams: Row hammer with crosshair,'' in \emph{Proceedings of the Second International Symposium on Memory Systems}, ser. MEMSYS '16.\hskip 1em plus 0.5em minus 0.4em\relax New York, NY, USA: Association for Computing Machinery, 2016, p. 471–476.
\BIBentrySTDinterwordspacing

\bibitem{winefs}
\BIBentryALTinterwordspacing
R.~Kadekodi, S.~Kadekodi, S.~Ponnapalli, H.~Shirwadkar, G.~R. Ganger, A.~Kolli, and V.~Chidambaram, ``Winefs: A hugepage-aware file system for persistent memory that ages gracefully,'' in \emph{Proceedings of the ACM SIGOPS 28th Symposium on Operating Systems Principles}, ser. SOSP '21.\hskip 1em plus 0.5em minus 0.4em\relax New York, NY, USA: Association for Computing Machinery, 2021, p. 804–818.
\BIBentrySTDinterwordspacing

\bibitem{pmemcheck}
T.~Kapela, ``An introduction to pmemcheck,'' 2015, \url{https://pmem.io/blog/2015/07/an-introduction-to-pmemcheck-part-1-basics/}.

\bibitem{8615701}
J.~Kim, M.~Patel, H.~Hassan, and O.~Mutlu, ``Solar-dram: Reducing dram access latency by exploiting the variation in local bitlines,'' in \emph{2018 IEEE 36th International Conference on Computer Design (ICCD)}, 2018, pp. 282--291.

\bibitem{linefs}
\BIBentryALTinterwordspacing
J.~Kim, I.~Jang, W.~Reda, J.~Im, M.~Canini, D.~Kosti\'{c}, Y.~Kwon, S.~Peter, and E.~Witchel, ``Linefs: Efficient smartnic offload of a distributed file system with pipeline parallelism,'' in \emph{Proceedings of the ACM SIGOPS 28th Symposium on Operating Systems Principles}, ser. SOSP '21.\hskip 1em plus 0.5em minus 0.4em\relax New York, NY, USA: Association for Computing Machinery, 2021, p. 756–771.
\BIBentrySTDinterwordspacing

\bibitem{6853210}
Y.~Kim, R.~Daly, J.~Kim, C.~Fallin, J.~H. Lee, D.~Lee, C.~Wilkerson, K.~Lai, and O.~Mutlu, ``Flipping bits in memory without accessing them: An experimental study of dram disturbance errors,'' in \emph{2014 ACM/IEEE 41st International Symposium on Computer Architecture (ISCA)}, 2014, pp. 361--372.

\bibitem{KlimisICSE24}
V.~Klimis, A.~F. Donaldson, V.~Vafeiadis, J.~Wickerson, and A.~Raad, ``Challenges in empirically testing memory persistency models,'' in \emph{Proceedings of the 2024 ACM/IEEE 44th International Conference on Software Engineering: New Ideas and Emerging Results}, ser. ICSE-NIER'24, 2024, p. 82–86.

\bibitem{strata}
\BIBentryALTinterwordspacing
Y.~Kwon, H.~Fingler, T.~Hunt, S.~Peter, E.~Witchel, and T.~Anderson, ``Strata: A cross media file system,'' in \emph{Proceedings of the 26th Symposium on Operating Systems Principles}, ser. SOSP '17.\hskip 1em plus 0.5em minus 0.4em\relax New York, NY, USA: Association for Computing Machinery, 2017, p. 460–477.
\BIBentrySTDinterwordspacing

\bibitem{10.1145/3084464}
\BIBentryALTinterwordspacing
D.~Lee, S.~Khan, L.~Subramanian, S.~Ghose, R.~Ausavarungnirun, G.~Pekhimenko, V.~Seshadri, and O.~Mutlu, ``Design-induced latency variation in modern dram chips: Characterization, analysis, and latency reduction mechanisms,'' \emph{Proc. ACM Meas. Anal. Comput. Syst.}, vol.~1, no.~1, jun 2017.
\BIBentrySTDinterwordspacing

\bibitem{pmfuzz}
\BIBentryALTinterwordspacing
S.~Liu, S.~Mahar, B.~Ray, and S.~M. Khan, ``Pmfuzz: test case generation for persistent memory programs,'' in \emph{{ASPLOS} '21: 26th {ACM} International Conference on Architectural Support for Programming Languages and Operating Systems, Virtual Event, USA, April 19-23, 2021}, T.~Sherwood, E.~D. Berger, and C.~Kozyrakis, Eds.\hskip 1em plus 0.5em minus 0.4em\relax {ACM}, 2021, pp. 487--502.
\BIBentrySTDinterwordspacing

\bibitem{xfdetector}
\BIBentryALTinterwordspacing
S.~Liu, K.~Seemakhupt, Y.~Wei, T.~F. Wenisch, A.~Kolli, and S.~M. Khan, ``Cross-failure bug detection in persistent memory programs,'' in \emph{{ASPLOS} '20: Architectural Support for Programming Languages and Operating Systems, Lausanne, Switzerland, March 16-20, 2020}, J.~R. Larus, L.~Ceze, and K.~Strauss, Eds.\hskip 1em plus 0.5em minus 0.4em\relax {ACM}, 2020, pp. 1187--1202.
\BIBentrySTDinterwordspacing

\bibitem{pmtest}
\BIBentryALTinterwordspacing
S.~Liu, Y.~Wei, J.~Zhao, A.~Kolli, and S.~M. Khan, ``Pmtest: {A} fast and flexible testing framework for persistent memory programs,'' in \emph{Proceedings of the Twenty-Fourth International Conference on Architectural Support for Programming Languages and Operating Systems, {ASPLOS} 2019, Providence, RI, USA, April 13-17, 2019}, I.~Bahar, M.~Herlihy, E.~Witchel, and A.~R. Lebeck, Eds.\hskip 1em plus 0.5em minus 0.4em\relax {ACM}, 2019, pp. 411--425.
\BIBentrySTDinterwordspacing

\bibitem{lustig+17}
\BIBentryALTinterwordspacing
D.~Lustig, A.~Wright, A.~Papakonstantinou, and O.~Giroux, ``Automated synthesis of comprehensive memory model litmus test suites,'' in \emph{Proceedings of the Twenty-Second International Conference on Architectural Support for Programming Languages and Operating Systems, {ASPLOS} 2017, Xi'an, China, April 8-12, 2017}, Y.~Chen, O.~Temam, and J.~Carter, Eds.\hskip 1em plus 0.5em minus 0.4em\relax {ACM}, 2017, pp. 661--675.
\BIBentrySTDinterwordspacing

\bibitem{sstore}
\BIBentryALTinterwordspacing
J.~Meehan, N.~Tatbul, S.~Zdonik, C.~Aslantas, U.~Cetintemel, J.~Du, T.~Kraska, S.~Madden, D.~Maier, A.~Pavlo, M.~Stonebraker, K.~Tufte, and H.~Wang, ``S-store: Streaming meets transaction processing,'' \emph{Proc. VLDB Endow.}, vol.~8, no.~13, p. 2134–2145, sep 2015.
\BIBentrySTDinterwordspacing

\bibitem{DRAMA}
\BIBentryALTinterwordspacing
P.~Pessl, D.~Gruss, C.~Maurice, M.~Schwarz, and S.~Mangard, ``{DRAMA}: Exploiting {DRAM} addressing for {Cross-CPU} attacks,'' in \emph{25th USENIX Security Symposium (USENIX Security 16)}.\hskip 1em plus 0.5em minus 0.4em\relax Austin, TX: USENIX Association, Aug. 2016, pp. 565--581.
\BIBentrySTDinterwordspacing

\bibitem{RaadMV22}
\BIBentryALTinterwordspacing
A.~Raad, L.~Maranget, and V.~Vafeiadis, ``Extending intel-x86 consistency and persistency: formalising the semantics of intel-x86 memory types and non-temporal stores,'' \emph{Proc. {ACM} Program. Lang.}, vol.~6, no. {POPL}, pp. 1--31, 2022.
\BIBentrySTDinterwordspacing

\bibitem{RaadWNV20}
\BIBentryALTinterwordspacing
A.~Raad, J.~Wickerson, G.~Neiger, and V.~Vafeiadis, ``Persistency semantics of the intel-x86 architecture,'' \emph{Proc. {ACM} Program. Lang.}, vol.~4, no. {POPL}, pp. 11:1--11:31, 2020.
\BIBentrySTDinterwordspacing

\bibitem{RaadWV19}
\BIBentryALTinterwordspacing
A.~Raad, J.~Wickerson, and V.~Vafeiadis, ``Weak persistency semantics from the ground up: formalising the persistency semantics of armv8 and transactional models,'' \emph{Proc. {ACM} Program. Lang.}, vol.~3, no. {OOPSLA}, pp. 135:1--135:27, 2019.
\BIBentrySTDinterwordspacing

\bibitem{Rudoff}
\BIBentryALTinterwordspacing
A.~Rudoff, ``Persistent memory programming,'' 2017, \url{https://www.usenix.org/system/files/login/articles/login_summer17_07_rudoff.pdf}.
\BIBentrySTDinterwordspacing

\bibitem{deep_persistence}
\BIBentryALTinterwordspacing
A.~J. Rushing, ``Persistent memory cleaning,'' May 2020, united States Patent Application 20200142774.
\BIBentrySTDinterwordspacing

\bibitem{samsungpressrelease}
\BIBentryALTinterwordspacing
{Samsung Electronics}, ``Samsung electronics unveils far-reaching, next-generation memory solutions at flash memory summit 2022,'' August 2022.
\BIBentrySTDinterwordspacing

\bibitem{SarkarSAMW11}
\BIBentryALTinterwordspacing
S.~Sarkar, P.~Sewell, J.~Alglave, L.~Maranget, and D.~Williams, ``Understanding {POWER} multiprocessors,'' in \emph{Proceedings of the 32nd {ACM} {SIGPLAN} Conference on Programming Language Design and Implementation, {PLDI} 2011, San Jose, CA, USA, June 4-8, 2011}, M.~W. Hall and D.~A. Padua, Eds.\hskip 1em plus 0.5em minus 0.4em\relax {ACM}, 2011, pp. 175--186.
\BIBentrySTDinterwordspacing

\bibitem{Scargall2020}
\BIBentryALTinterwordspacing
S.~Scargall, \emph{Persistent Memory Architecture}.\hskip 1em plus 0.5em minus 0.4em\relax Berkeley, CA: Apress, 2020, pp. 11--30.
\BIBentrySTDinterwordspacing

\bibitem{olive}
\BIBentryALTinterwordspacing
S.~Setty, C.~Su, J.~R. Lorch, L.~Zhou, H.~Chen, P.~Patel, and J.~Ren, ``Realizing the {Fault-Tolerance} promise of cloud storage using locks with intent,'' in \emph{12th USENIX Symposium on Operating Systems Design and Implementation (OSDI 16)}.\hskip 1em plus 0.5em minus 0.4em\relax Savannah, GA: USENIX Association, Nov. 2016, pp. 501--516.
\BIBentrySTDinterwordspacing

\bibitem{SewellSONM10}
\BIBentryALTinterwordspacing
P.~Sewell, S.~Sarkar, S.~Owens, F.~Z. Nardelli, and M.~O. Myreen, ``x86-tso: a rigorous and usable programmer's model for x86 multiprocessors,'' \emph{Commun. {ACM}}, vol.~53, no.~7, pp. 89--97, 2010.
\BIBentrySTDinterwordspacing

\bibitem{SimnerAPPGS22}
\BIBentryALTinterwordspacing
B.~Simner, A.~Armstrong, J.~Pichon{-}Pharabod, C.~Pulte, R.~Grisenthwaite, and P.~Sewell, ``Relaxed virtual memory in armv8-a,'' in \emph{Programming Languages and Systems - 31st European Symposium on Programming, {ESOP} 2022, Held as Part of the European Joint Conferences on Theory and Practice of Software, {ETAPS} 2022, Munich, Germany, April 2-7, 2022, Proceedings}, ser. Lecture Notes in Computer Science, I.~Sergey, Ed., vol. 13240.\hskip 1em plus 0.5em minus 0.4em\relax Springer, 2022, pp. 143--173.
\BIBentrySTDinterwordspacing

\bibitem{10.1007/978-3-030-00470-5_3}
A.~Tatar, C.~Giuffrida, H.~Bos, and K.~Razavi, ``Defeating software mitigations against rowhammer: A surgical precision hammer,'' in \emph{Research in Attacks, Intrusions, and Defenses}, M.~Bailey, T.~Holz, M.~Stamatogiannakis, and S.~Ioannidis, Eds.\hskip 1em plus 0.5em minus 0.4em\relax Cham: Springer International Publishing, 2018, pp. 47--66.

\bibitem{kafka}
\BIBentryALTinterwordspacing
G.~Wang, L.~Chen, A.~Dikshit, J.~Gustafson, B.~Chen, M.~J. Sax, J.~Roesler, S.~Blee-Goldman, B.~Cadonna, A.~Mehta, V.~Madan, and J.~Rao, \emph{Consistency and Completeness: Rethinking Distributed Stream Processing in Apache Kafka}.\hskip 1em plus 0.5em minus 0.4em\relax New York, NY, USA: Association for Computing Machinery, 2021, p. 2602–2613.
\BIBentrySTDinterwordspacing

\bibitem{wang22}
\BIBentryALTinterwordspacing
W.~Wang, ``Architectural support for persistent memory,'' in \emph{Workshop on Novel Architecture and Novel Design Automation (NANDA)}, 2022.
\BIBentrySTDinterwordspacing

\bibitem{197231}
\BIBentryALTinterwordspacing
Y.~Xiao, X.~Zhang, Y.~Zhang, and R.~Teodorescu, ``One bit flips, one cloud flops: {Cross-VM} row hammer attacks and privilege escalation,'' in \emph{25th USENIX Security Symposium (USENIX Security 16)}.\hskip 1em plus 0.5em minus 0.4em\relax Austin, TX: USENIX Association, Aug. 2016, pp. 19--35.
\BIBentrySTDinterwordspacing

\bibitem{nova}
\BIBentryALTinterwordspacing
J.~Xu and S.~Swanson, ``{NOVA:} {A} log-structured file system for hybrid volatile/non-volatile main memories,'' in \emph{14th {USENIX} Conference on File and Storage Technologies, {FAST} 2016, Santa Clara, CA, USA, February 22-25, 2016}, A.~D. Brown and F.~I. Popovici, Eds.\hskip 1em plus 0.5em minus 0.4em\relax {USENIX} Association, 2016, pp. 323--338.
\BIBentrySTDinterwordspacing

\bibitem{beldi}
\BIBentryALTinterwordspacing
H.~Zhang, A.~Cardoza, P.~B. Chen, S.~Angel, and V.~Liu, ``Fault-tolerant and transactional stateful serverless workflows,'' in \emph{14th USENIX Symposium on Operating Systems Design and Implementation (OSDI 20)}.\hskip 1em plus 0.5em minus 0.4em\relax USENIX Association, Nov. 2020, pp. 1187--1204.
\BIBentrySTDinterwordspacing

\bibitem{octopus}
\BIBentryALTinterwordspacing
B.~Zhu, Y.~Chen, Q.~Wang, Y.~Lu, and J.~Shu, ``Octopus\({}^{\mbox{+}}\): An rdma-enabled distributed persistent memory file system,'' \emph{{ACM} Trans. Storage}, vol.~17, no.~3, pp. 19:1--19:25, 2021.
\BIBentrySTDinterwordspacing

\end{thebibliography}

\end{document}